\documentclass[longauth,traditabstract,letter]{aa} 
\usepackage{graphicx}
\usepackage{txfonts}
\usepackage{natbib}
\usepackage{float}
\begin{document}
\title{\emph{Herschel}-HIFI detections of hydrides towards AFGL 2591\thanks{\emph{Herschel} is an ESA space observatory with science instruments provided by European-led Principal Investigator consortia and with important participation from NASA.}}

\subtitle{Envelope emission versus tenuous cloud absorption}

\author{
     S.~Bruderer\inst{\ref{inst3}}
\and A.O.~Benz\inst{\ref{inst3}}
\and E.F.~van~Dishoeck\inst{\ref{inst1},\ref{inst2}}
\and M. Melchior\inst{\ref{inst47}}
\and S.D.~Doty\inst{\ref{inst19}}
\and F.~van der Tak\inst{\ref{inst10},\ref{inst11}}
\and P.~St\"auber\inst{\ref{inst3}}
\and S.F.~Wampfler\inst{\ref{inst3}}
\and C.~Dedes\inst{\ref{inst3}}
\and U.A.~Y{\i}ld{\i}z\inst{\ref{inst1}}
\and L.~Pagani\inst{\ref{inst20}}
\and T.~Giannini\inst{\ref{inst23}}
\and Th.~de~Graauw\inst{\ref{inst52}}
\and N.~Whyborn\inst{\ref{inst52}}
\and D.~Teyssier\inst{\ref{inst49}}
\and W.~Jellema\inst{\ref{inst10}}
\and R.~Shipman\inst{\ref{inst10}}
\and R.~Schieder\inst{\ref{inst43}}
\and N.~Honingh\inst{\ref{inst43}}
\and E.~Caux\inst{\ref{inst50},\ref{inst51}}
\and W.~B\"achtold\inst{\ref{inst48}}
\and A.~Csillaghy\inst{\ref{inst47}}
\and C.~Monstein\inst{\ref{inst3}}
\and R.~Bachiller\inst{\ref{inst12}}
\and A.~Baudry\inst{\ref{inst6}}
\and M.~Benedettini\inst{\ref{inst13}}
\and E.~Bergin\inst{\ref{inst14}}
\and P.~Bjerkeli\inst{\ref{inst9}}
\and G.A.~Blake\inst{\ref{inst15}}
\and S.~Bontemps\inst{\ref{inst6}}
\and J.~Braine\inst{\ref{inst6}}
\and P.~Caselli\inst{\ref{inst4},\ref{inst5}}
\and J.~Cernicharo\inst{\ref{inst16}}
\and C.~Codella\inst{\ref{inst5}}
\and F.~Daniel\inst{\ref{inst16}}
\and A.M.~di~Giorgio\inst{\ref{inst13}}
\and C.~Dominik\inst{\ref{inst17},\ref{inst18}}
\and P.~Encrenaz\inst{\ref{inst20}}
\and M.~Fich\inst{\ref{inst21}}
\and A.~Fuente\inst{\ref{inst22}}
\and J.R.~Goicoechea\inst{\ref{inst16}}
\and F.~Helmich\inst{\ref{inst10}}
\and G.J.~Herczeg\inst{\ref{inst2}}
\and F.~Herpin\inst{\ref{inst6}}
\and M.R.~Hogerheijde\inst{\ref{inst1}}
\and T.~Jacq\inst{\ref{inst5}}
\and D.~Johnstone\inst{\ref{inst7},\ref{inst8}}
\and J.K.~J{\o}rgensen\inst{\ref{inst24}}
\and L.E.~Kristensen\inst{\ref{inst1}}
\and B.~Larsson\inst{\ref{inst25}}
\and D.~Lis\inst{\ref{inst26}}
\and R.~Liseau\inst{\ref{inst9}}
\and M.~Marseille\inst{\ref{inst10}}
\and C.~M$^{\textrm c}$Coey\inst{\ref{inst21},\ref{inst27}}
\and G.~Melnick\inst{\ref{inst28}}
\and D.~Neufeld\inst{\ref{inst29}}
\and B.~Nisini\inst{\ref{inst23}}
\and M.~Olberg\inst{\ref{inst9}}
\and B.~Parise\inst{\ref{inst30}}
\and J.C.~Pearson\inst{\ref{inst31}}
\and R.~Plume\inst{\ref{inst32}}
\and C.~Risacher\inst{\ref{inst10}}
\and J.~Santiago-Garc\'{i}a\inst{\ref{inst33}}
\and P.~Saraceno\inst{\ref{inst13}}
\and R.~Shipman\inst{\ref{inst10}}
\and M.~Tafalla\inst{\ref{inst12}}
\and T.A.~van~Kempen\inst{\ref{inst28}}
\and R.~Visser\inst{\ref{inst1}}
\and F.~Wyrowski\inst{\ref{inst30}}}

\institute{
Institute of Astronomy, ETH Zurich, 8093 Zurich, Switzerland\label{inst3}
\and
Leiden Observatory, Leiden University, PO Box 9513, 2300 RA Leiden, The Netherlands\label{inst1}
\and
Max Planck Institut f\"{u}r Extraterrestrische Physik, Giessenbachstrasse 1, 85748 Garching, Germany\label{inst2}
\and
Institut f\"ur 4D-Technologien, FHNW, 5210 Windisch, Switzerland\label{inst47}
\and
Department of Physics and Astronomy, Denison University, Granville, OH, 43023, USA\label{inst19}
\and
SRON Netherlands Institute for Space Research, PO Box 800, 9700 AV, Groningen, The Netherlands\label{inst10}
\and
Kapteyn Astronomical Institute, University of Groningen, PO Box 800, 9700 AV, Groningen, The Netherlands\label{inst11}
\and
LERMA and UMR 8112 du CNRS, Observatoire de Paris, 61 Av. de l'Observatoire, 75014 Paris, France\label{inst20}
\and
INAF - Osservatorio Astronomico di Roma, 00040 Monte Porzio catone, Italy\label{inst23}
\and
Atacama Large Millimeter/Submillimeter Array, Joint ALMA Office, Santiago, Chile\label{inst52}
\and
European Space Astronomy Centre, ESA, P.O. Box 78, E-28691 Villanueva de la Ca\~nada, Madrid\label{inst49}
\and
KOSMA, I. Physik. Institut, Universit\"{a}t zu K\"{o}ln, Z\"{u}lpicher Str. 77, D 50937 K\"{o}ln, Germany\label{inst43}
\and 
Centre d'Etude Spatiale des Rayonnements, Universit\'e de Toulouse [UPS], 31062 Toulouse Cedex 9, France\label{inst50}
\and 
CNRS/INSU, UMR 5187, 9 avenue du Colonel Roche, 31028 Toulouse Cedex 4, France\label{inst51}
\and 
Laboratory for Electromagnetic Fields ad Microwave Electronics, ETH Zurich, 8092 Zurich, Switzerland\label{inst48}
\and
Observatorio Astron\'{o}mico Nacional (IGN), Calle Alfonso XII,3. 28014, Madrid, Spain\label{inst12}
\and
Universit\'{e} de Bordeaux, Laboratoire d'Astrophysique de Bordeaux, France; CNRS/INSU, UMR 5804, Floirac, France\label{inst6}
\and
INAF - Istituto di Fisica dello Spazio Interplanetario, Area di Ricerca di Tor Vergata, via Fosso del Cavaliere 100, 00133 Roma, Italy\label{inst13}
\and
Department of Astronomy, The University of Michigan, 500 Church Street, Ann Arbor, MI 48109-1042, USA\label{inst14}
\and
Department of Radio and Space Science, Chalmers University of Technology, Onsala Space Observatory, 439 92 Onsala, Sweden\label{inst9}
\and
California Institute of Technology, Division of Geological and Planetary Sciences, MS 150-21, Pasadena, CA 91125, USA\label{inst15}
\and
School of Physics and Astronomy, University of Leeds, Leeds LS2 9JT, UK\label{inst4}
\and
INAF - Osservatorio Astrofisico di Arcetri, Largo E. Fermi 5, 50125 Firenze, Italy\label{inst5}
\and
Centro de Astrobiolog\'{\i}a. Departamento de Astrof\'{\i}sica. CSIC-INTA. Carretera de Ajalvir, Km 4, Torrej\'{o}n de Ardoz. 28850, Madrid, Spain.\label{inst16}
\and
Astronomical Institute Anton Pannekoek, University of Amsterdam, Kruislaan 403, 1098 SJ Amsterdam, The Netherlands\label{inst17}
\and
Department of Astrophysics/IMAPP, Radboud University Nijmegen, P.O. Box 9010, 6500 GL Nijmegen, The Netherlands\label{inst18}
\and
University of Waterloo, Department of Physics and Astronomy, Waterloo, Ontario, Canada\label{inst21}
\and
Observatorio Astron\'{o}mico Nacional, Apartado 112, 28803 Alcal\'{a} de Henares, Spain\label{inst22}
\and
National Research Council Canada, Herzberg Institute of Astrophysics, 5071 West Saanich Road, Victoria, BC V9E 2E7, Canada\label{inst7}
\and
Department of Physics and Astronomy, University of Victoria, Victoria, BC V8P 1A1, Canada\label{inst8}
\and
Centre for Star and Planet Formation, Natural History Museum of Denmark, University of Copenhagen,
{\O}ster Voldgade 5-7, DK-1350 Copenhagen K., Denmark\label{inst24}
\and
Department of Astronomy, Stockholm University, AlbaNova, 106 91 Stockholm, Sweden\label{inst25}
\and
California Institute of Technology, Cahill Center for Astronomy and Astrophysics, MS 301-17, Pasadena, CA 91125, USA\label{inst26}
\and
the University of Western Ontario, Department of Physics and Astronomy, London, Ontario, N6A 3K7, Canada\label{inst27}
\and
Harvard-Smithsonian Center for Astrophysics, 60 Garden Street, MS 42, Cambridge, MA 02138, USA\label{inst28}
\and
Department of Physics and Astronomy, Johns Hopkins University, 3400 North Charles Street, Baltimore, MD 21218, USA\label{inst29}
\and
Max-Planck-Institut f\"{u}r Radioastronomie, Auf dem H\"{u}gel 69, 53121 Bonn, Germany\label{inst30}
\and
Jet Propulsion Laboratory, California Institute of Technology, Pasadena, CA 91109, USA\label{inst31}
\and
Department of Physics and Astronomy, University of Calgary, Calgary, T2N 1N4, AB, Canada\label{inst32}
\and
Instituto de Radioastronom\'{i}a Milim\'{e}trica (IRAM), Avenida Divina Pastora 7, N\'{u}cleo Central, E-18012 Granada, Spain\label{inst33}
}

\date{Received \today; accepted}

\titlerunning{Hydrides towards AFGL 2591}
\authorrunning{S. Bruderer et al.}

\offprints{Simon Bruderer,\\ \email{simonbr@astro.phys.ethz.ch}}

\abstract{The Heterodyne Instrument for the Far Infrared (HIFI) onboard the \emph{Herschel} Space Observatory allows the first observations of light diatomic molecules at high spectral resolution and in multiple transitions. Here, we report deep integrations using HIFI in different lines of hydrides towards the high-mass star forming region AFGL 2591. Detected are CH, CH$^+$, NH, OH$^+$, H$_2$O$^+$, while NH$^+$ and SH$^+$ have not been detected. All molecules except for CH and CH$^+$ are seen in absorption with low excitation temperatures and at velocities different from the systemic velocity of the protostellar envelope. Surprisingly, the CH($J_{F,P} = 3/2_{2,-} - 1/2_{1,+}$) and CH$^+$($J=1-0$, $J=2-1$) lines are detected in emission at the systemic velocity. We can assign the absorption features to a foreground cloud and an outflow lobe, while the CH and CH$^+$ emission stems from the envelope. The observed abundance and excitation of CH and CH$^+$ can be explained in the scenario of FUV irradiated outflow walls, where a cavity etched out by the outflow allows protostellar FUV photons to irradiate and heat the envelope at larger distances driving the chemical reactions that produce these molecules.}

\maketitle


\section{Introduction} \label{sec:intro}

The simplest constituents of interstellar chemistry, light diatomic hydrides and their ions (XH and XH$^+$ with X$=$O, H, C and N), have so far not been studied thoroughly. Since their rotational ground state lines lie mostly outside atmospheric windows, they cannot be observed from ground. However, they are key-species of the chemical network. For example, most water at high temperature ($T>250$ K) is formed by the reaction of OH with H$_2$. Moreover, once their chemistry and excitation is well understood, they may be valuable tracers of warm and partly ionized gas (e.g. \citealt{Sternberg95,Bruderer10}). This is of particular interest in the context of star formation, as the radiation of the forming star may heat and ionize the envelope giving birth to it.

Chemistry and excitation of hydrides is characterized by high activation energies and high critical densities. For example, the only formation route of CH$^+$ is via the highly endothermic reaction C$^+$ + H$_2$ + 4640 K  $\rightarrow$ CH$^+$ + H. The critical density\footnote{The density, at which collisional deexcitation is comparable to spontaneous decay.} for many hydrides is very high. For example the CH$^+$($J=1-0$) transition has a critical density of $\sim 5 \times 10^7$ cm$^{-3}$ for collisions with H$_2$ at 500 K (\citealt{Hammami09}\footnote{Scaled for the reduced masses of the H$_2$-CH$^+$ system.}). This very reactive ion may however be destroyed rather than excited in collisions with H$_2$ (\citealt{Black98}). Thus, the excitation of the ions studied here may be governed by the initial state in which they are formed rather than by equilibrium with collisions or the radiation field. The situation of considerable abundance only at high temperature ($T \gg 100$ K), high critical density and quick destruction in collisions with H$_2$ is similar to CO$^+$, studied by \citet{Staeuber09}. Finally, if a molecule is abundant, but not excited, it may still act as an absorber of a background continuum source and shows up as an absorption line.

The Heterodyne Instrument for the Far Infrared (HIFI; \citealt{deGraauw10}) onboard the \emph{Herschel} Space Observatory allows for the first time to carry out spectrally resolved observations of different transitions of hydrides and thus to disentangle moving regions along the line of sight and to study fine/hyperfine structure components.

\object{AFGL 2591} is a well studied high-mass star forming region at a distance of $\sim1$ kpc (e.g. \citealt{vdTak99}). The source is relatively nearby and its geometry thus well known (e.g. \citealt{Preibisch03}). There is evidence for a cavity along the outflow that allows FUV photons to escape to longer distances and irradiate the outflow walls (\citealt{Bruderer09c,Bruderer09b}). In these high-density PDRs, the abundances of hydrides can be enhanced by several orders of magnitude. Chemical models of this scenario predict fluxes for the CH$^+$($J=1-0$) line of up to 16 K km s$^{-1}$ for AFGL 2591, while those from a spherical model are far below the detection limit of \emph{Herschel} (\citealt{Bruderer10}).

\section{Observations and data reduction} \label{sec:obs}

HIFI-\emph{Herschel} observations of AFGL 2591 (\mbox{$\alpha_{2000}=20^{\rm h}29^{\rm m}24\fs87$}, \mbox{$\delta_{2000}=+40\degr 11\arcmin 19.5\arcsec$}) have been carried out as part of the Water In Star forming regions with \emph{Herschel} (WISH; van Dishoeck et al., in prep.) guaranteed time key program. Target lines are given in Table \ref{tab:lines}. Molecular parameters are obtained from the CDMS (\citealt{Mueller05}), except for NH$^+$ (\citealt{Huebers09}) and H$_2$O$^+$ (\citealt{Muertz98}). Dual beam switch mode with fast chopping was used. To disentangle lines from the upper and lower side band, the local oscillator frequency was slightly shifted after half of the integration time. Data were converted to $T_{\rm mb}$ using a main beam efficiency of 0.74. Data from the WBS spectrometer are used and the H and V polarizations have been averaged after visual inspection. Table \ref{tab:lines} gives the noise level of the observation $T_{\rm rms}$ for the nominal resolution of the WBS spectrometer (1.1 MHz). Data were reduced with the \emph{Herschel} interactive processing environment (HIPE) package version 2.9.

\section{Results} \label{sec:result}

Spectra of detected target lines are given in Fig. \ref{fig:spectra}, those without detection in the Appendix \ref{sec:nondetect}. We have detected the ground state lines of CH, CH$^+$, OH$^+$, H$_2$O$^+$ and NH. The lines of OH$^+$, H$_2$O$^+$ and NH are seen in absorption without clear signs of emission on top of the absorption. The CH line is almost purely in emission. The CH$^+$($J=1-0$) line is in absorption, with emission at the systemic velocity of the envelope. The CH$^+$($J=2-1$) line is in emission. Except for CH$^+$($J=2-1$), no line between excited states is detected. Not detected are SH$^+$ and NH$^+$. While SH$^+$ has been seen in other regions (\citealt{Benz10,Menten10}), interstellar NH$^+$ has not yet been found.

By their velocity shift, the absorption lines of CH, CH$^+$, NH, OH$^+$ and H$_2$O$^+$ can be roughly classified as (i) a component around v$_{\rm lsr} \sim 0$ km s$^{-1}$, red-shifted with respect to the systemic velocity of the envelope at v$_{\rm sys} = -5.5$ km s$^{-1}$ (\citealt{vdTak99}) and (ii) a blue-shifted component at $\sim -11$ km s$^{-1}$ (NH) or $\sim -16$ km s$^{-1}$ (OH$^+$, H$_2$O$^+$ and CH$^+$).

The velocity integrated emission or absorption is given in Table \ref{tab:lines}. The absorption per frequency is $\tau=-\log(T_{\rm line}/T_{\rm cont})$, with the continuum intensity $T_{\rm cont}$ in a single side band. The velocity integrated emission and absorption are related to the molecular column density of the upper or lower level $N^{\rm upper}$ and $N^{\rm lower}$ by 
\begin{equation}
N^{\rm upper}= \frac{8 \pi k}{h c^3} \frac{\nu_{ul}^2}{A_{ul}} \int T d{\rm v}  \ \ \textrm{ and } \ \ N^{\rm lower} = \frac{8 \pi}{c^3} \frac{\nu_{ul}^3 g_l}{A_{ul} g_u}  \int \tau d{\rm v}  \ ,
\end{equation}
with the Einstein-A coefficient $A_{ul}$ and the frequency $\nu_{ul}$ of the transition and the statistical weights  $g_u$ and $g_l$ of the levels. These equations assume optically thin emission or absorption without re-emission. For Table \ref{tab:lines}, contributions of different fine/hyperfine components are summed up, weighted by their statistical weights. Detected lines together with the upper limits from higher levels yield upper limits on the excitation temperature and the column density, assuming the excitation temperature of all lines to be the same. For CH, we obtain $T_{\rm ex} < 22.6$ K and $N({\rm CH}) < 1.6 \times 10^{13}$ cm$^{-2}$, OH$^+$ yields $T_{\rm ex} < 13.2$ K and $N({\rm OH}^+) < 1.2 \times 10^{14}$ cm$^{-2}$ and H$_2$O$^+$ gives $T_{\rm ex} < 18.5$ K and $N({\rm H}_2{\rm O}^+) < 2.3 \times 10^{13}$ cm$^{-2}$. We refrain from deriving column densities for CH$^+$ using this method, as the molecule shows clear signs of both emission and absorption.

\subsection{Modeling the line spectra} \label{sec:specfit}

\begin{figure*}[tb]
  \includegraphics[width=\hsize]{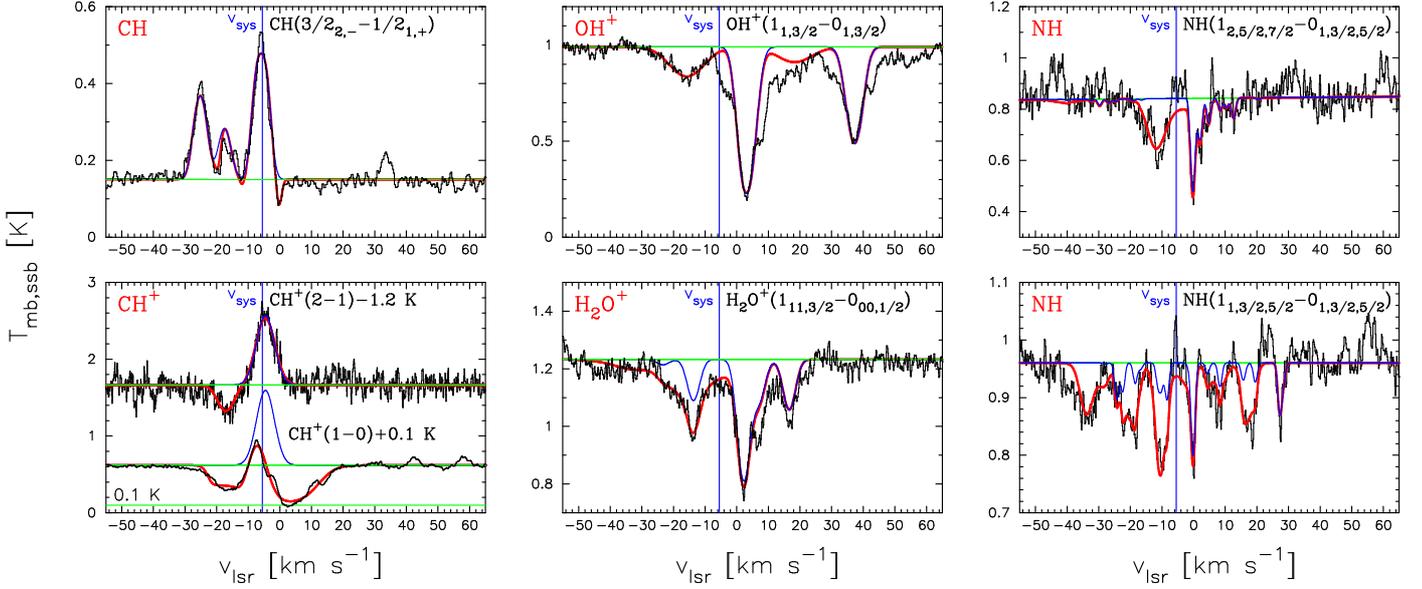}
  \caption{HIFI-Spectra of hydrides observed towards AFGL 2591. The green line represents the continuum level used for the calculation of the velocity integrated emission or absorption. The vertical blue line indicates the systemic velocity of -5.5 km s$^{-1}$ of AFGL 2591. The red line shows a fitted model spectrum discussed in Section \ref{sec:specfit}, with parameters given in Table \ref{tab:fit}. The blue line indicates a model spectrum with only one of the components, showing the relative intensity of the different hyperfine components. The spectra are corrected for the beam efficiency and the continuum level is given for one sideband (SSB).}\label{fig:spectra}
\end{figure*}

The line profile is modeled using a two-layer model (red and blue) solving the radiative transfer equation along the line of sight (Appendix \ref{sec:rtsolution}). Slab geometry is assumed and the absorbing/emitting foreground layers cover the whole background continuum. Free parameters per component are the column density, the excitation temperature and the line width and position. The intrinsic line profile is assumed to be a Gaussian. The upper limits of the OH$^+$ and H$_2$O$^+$ excitation temperature are lower than the energy of the upper level and for such low temperatures, the absorption lines do not depend much on the excitation temperature. We thus fix $T_{\rm ex}$ at \mbox{3 K}. Parameters of the best-fit model are given in Table \ref{tab:fit}. The model spectra are overlayed on the observed spectra in Figure \ref{fig:spectra}. The figure also gives only one component, showing the necessity of a second component. A good fit is obtained for the CH, NH and  H$_2$O$^+$ lines. For CH and OH$^+$, further components would improve the fit. For OH$^+$, an asymmetric line would presumably be needed to fit observations. However, for the conclusions drawn here, we concentrate on the strongest components. A third component has only been added to model the emission peak of CH$^+$ at the systemic velocity of the envelope. The CH column densities derived with this method are larger compared to the calculation assuming optically thin lines. This is because the fitting of the different fine structure components requires a peak optical depth of $\tau \sim 2$.

\begin{table*}[tb]
\caption{Observed lines in AFGL 2591.\label{tab:lines}}
\scriptsize
\centering
\begin{tabular}{lllc|cccc|ccc|rr}
\hline
\hline
\multicolumn{4}{l|}{Species/Transition/Electronic state/Line components$^f$}      & Frequency    &  $E_{u}$ & $A_{ul}$   &  $T_{\rm obs}^e$ &  $\int T d$v & $\int \tau d$v  & $T_{\rm rms}$ &  $N^{\rm upper}({\rm X})$ &  $N^{\rm lower}({\rm X})$ \\
   &                   &       & & [GHz]        &  [K]     & [s$^{-1}$] &  [min] & [K km s$^{-1}$] & [km s$^{-1}$]   & [mK]     &        [cm$^{-2}$]    & [cm$^{-2}$]         \\
\hline
CH &$J_{F,P}=3/2_{2,-}-1/2_{1,+}$                        & $^{2}\Pi$                & 3       &  536.7611    &  25.76   &  6.4(-4) & 4.2& 3.46$\pm$0.04$^a$& \ldots             &    16        &    3.0(12)       &      \ldots     \\
CH &$J_{F,P}=5/2_{3,+}-3/2_{2,-}$                        & $^{2}\Pi$                & 3       & 1661.1074    & 105.48   &  3.8(-2) &20.2&  $<0.93$          & $<0.34^{b,c}$      &   220        &  $<1.3(11)$      &  $<2.6(12)$     \\
CH$^+$ &$J=1-0$                                          & $^{1}\Sigma^+$           & 1       &  835.1375    &  40.08   &  6.4(-2) &40.4&  $0.91\pm0.03$$^d$& $32.63\pm0.06$$^d$ &   16         &    1.9(11)$^d$   &      9.3(13)$^d$\\
CH$^+$ &$J=2-1$                                          & $^{1}\Sigma^+$           & 1       & 1669.2813    & 120.19   &  6.1(-2) &40.4&     $3.7\pm0.2$   & \ldots             &   115        &    3.3(11)       &      \ldots     \\
o-H$_2$O$^+$ &$N_{K_aK_b,J}=1_{11,3/2}-0_{00,1/2}$       & $^{2}B_1$                & 5       & 1115.2041    &  53.51   &  3.1(-2) &39.2&   \ldots          & $7.65\pm0.05$      &    33        &      \ldots      &    1.6(13)      \\
o-H$_2$O$^+$ &$N_{K_aK_b,J}=2_{02,3/2}-1_{11,3/2}$       & $^{2}B_1$                & 7       &  746.26      &  89.33   &  5.5(-4) &40.6&  $<0.16$          & $<0.52$            &    25        &  $<2.6(11)$      &  $<3.0(13)$     \\
OH$^+$ &$N_{J,F}=1_{1,3/2}-0_{1,3/2}$                    & $^{3}\Sigma^-$           & 4       & 1033.1186    &  49.58   &  1.8(-2) &40.6&   \ldots          &$21.98\pm0.06$      &    32        &      \ldots      &    1.1(14)      \\
OH$^+$ &$N_{J,F}=2_{1,3/2}-1_{1,3/2}$                    & $^{3}\Sigma^-$           & 4       & 1892.2271    & 140.39   &  5.9(-2) &21.0&  $<0.83$          & $<0.28$            &   210        &  $<8.2(10)$      &  $<2.5(12)$     \\
NH &$N_{J,F_1,F}=1_{2,5/2,7/2}$-$0_{1,3/2,5/2}$          & $^{3}\Sigma^-$           & 21      &  974.4784    &  46.77   &  6.9(-3) & 7.0&   \ldots          & $3.90\pm0.09$      &    43        &      \ldots      &    2.9(13)      \\
NH &$N_{J,F_1,F}=1_{1,3/2,5/2}$-$0_{1,3/2,5/2}$          & $^{3}\Sigma^-$           & 21      &  999.9734    &  47.99   &  5.2(-3) &38.8&   \ldots          & $3.99\pm0.05$      &    29        &      \ldots      &    5.0(13)      \\
NH$^+$ &$J_{P}=3/2_{-}-1/2_{+}$                          & $^{2}\Pi$                & 5       & 1012.5400    &  48.59   &  5.4(-2) &38.8&  $<0.16$          & $<0.19$            &    30        &  $<5.9(9)$       &  $<1.7(10)$     \\
NH$^+$ &$J_{P}=3/2_{+}-1/2_{-}$                          & $^{2}\Pi$                & 7       & 1019.2107    &  48.91   &  5.5(-2) &40.6&  $<0.12$          & $<0.13$            &    22        &  $<4.4(9)$       &  $<1.2(10)$     \\
SH$^+$ &$N_{J,F}=1_{2,5/2}-0_{1,3/2}$                    & $^{3}\Sigma^-$           & 3       &  526.0479    &  25.25   &  9.6(-4) & 4.2&  $<0.083$         & $<0.53$            &    11        &  $<4.7(10)$      &  $<4.5(12)$     \\
\hline
\end{tabular}
\begin{flushleft}
\footnotesize{The frequency is given for the strongest fine/hyperfine component. The velocity integrated intensity or absorption is summed over all fine/hyperfine components. $A(B) \equiv A \times 10^B$. $^a 1 \sigma \equiv \sqrt{\Delta V \delta v} T_{\rm rms}$, with $\Delta V \sim 10$ km s$^{-1}$ (expected line width) and the channel width $\delta v$ (corresponding to 1.1 MHz). $^b 1 \sigma \equiv \sqrt{\Delta V \delta v} T_{\rm rms} / T_{\rm cont}$. $^c$Non-detections are given by the $3 \sigma$ value. $^d$Emitting or absorbing part only. $^e$Total integration time (on+off). $^f$Hyperfine or fine structure.}
\end{flushleft}
\end{table*} 

\section{Discussion} \label{sec:disc}

The blue-shifted component of OH$^+$, H$_2$O$^+$ and CH$^+$ suggests the association with the outflow lobe directed towards us. \citet{Mitchell89} have detected 4.7 $\mu$m ${}^{13}$CO($\nu=0-1$) absorption lines at v$_{\rm lsr} \sim -28$ km s$^{-1}$. They derive a ${}^{13}$CO column density of $1.1 \times 10^{17}$ cm$^{-2}$ and a temperature of 200 K. With better spectral resolution, \citet{vdTak99} find v$_{\rm lsr} \sim -21.5$ km s$^{-1}$, in better agreement with our observations. Their lines are broad and asymmetric, with strong wings to the blue, likely originating in winds. It is thus difficult to determine the line width and position. However, the ${}^{13}$CO lines reach the continuum at about $-45$ km s$^{-1}$ and are clearly broader than the OH$^+$, H$_2$O$^+$ and CH$^+$ lines. Except for CH$^+$, where the two-component model does not fit this component well, there is no evidence for asymmetric lines in this blue-shifted component. The component is not clearly detected in CH. The NH absorption is less shifted to the blue and narrower. The velocity shift of $\sim -10$ km s$^{-1}$ corresponds to a narrow absorption feature seen in ${}^{12}$CO($J=3-2$) JCMT observations (\citealt{vdTak99}). They conclude that the absorption must take place in cold and/or tenuous gas, as it is not found in ${}^{12}$CO($J=6-5$).

The red-shifted absorption at v$_{\rm lsr} = -0.2$ to $2.8$ km s$^{-1}$ is likely associated to a cold and extended foreground cloud, observed in CO (\citealt{Mitchell92,Hasegawa95}). The small line width in their spectra approximately agrees with the narrow absorption lines seen in CH, OH$^+$, H$_2$O$^+$ and NH. The broad absorption of CH$^+$ is surprising: since it is only seen in CH$^+$($J=1-0$), but not CH$^+$($J=2-1$), the excitation temperature must be lower than 8 K, used to fit the blue shifted component. This indicates either less dense or colder gas. The ionized species OH$^+$, H$_2$O$^+$ and CH$^+$ are more red-shifted and have broader lines, compared to the neutral CH and NH. 

The abundances of molecules in the absorption features are roughly estimated from the total column density ($N_{\rm H} = N({\rm H}) + 2N({\rm H}_2)$) obtained from ${}^{13}$CO column densities and presented in Table \ref{tab:fit}. We use a CO abundance of $n({\rm CO})/n({\rm H}_2) = 3.7 \times 10^{-4}$ (\citealt{Doty02}) and ${}^{12}$CO/${}^{13}$CO $\sim$ 77 (\citealt{Wilson94}). Since the red component is a low-density foreground cloud, the fractional abundances of CH, H$_2$O$^+$ and NH have been compared to a chemical model for a diffuse/translucent cloud (\citealt{LePetit04}) and found to agree within a factor of $\sim2$. The underprediction of CH$^+$ by a factor of more than 100 is well known for such models, but models of turbulent dissipation regions (\citealt{Godard09}) can explain fractional abundances reached here. OH$^+$ is underpredicted by a factor of 6 but the observed OH$^+$/H$_2$O$^+$ ratio of 7.3 is comparable to the value of $>4$ found by \citet{Gerin10} and explained by low density gas with a low molecular fraction.

The blue component is thought to arise in tenuous warm gas in the outflow lobes directed toward us. The observed OH$^+$ and H$_2$O$^+$ abundances are lower by factors of 7 and 16, respectively, but the OH$^+$/H$_2$O$^+$ ratio is still large, about 3. As shown in Figure 2 of \citet{Gerin10} such ratios are consistent with models with densities up to a few thousand cm$^{-3}$ and enhanced radiation fields, as expected in the outflow lobe. The fractional abundances of CH$^+$ and NH are similar to those found in the red component within a factor of $\sim 2$. More specific modeling of these components must await better determinations of their physical conditions.

Where does the CH and CH$^+$ emission emerge from? The line velocity suggests that the emission stems from the envelope of AFGL 2591. Abundances of CH and CH$^+$ are difficult to estimate from column densities, due to the strong gradient in density in the envelope and thus the changing excitation. A spherical power-law density profile of the envelope (\citealt{vdTak99}) yields a density of $n({\rm H}_2) \sim 10^6$ cm$^{-3}$ at scales of the \emph{Herschel} beam. Relative to the column density of $N_{\rm H} \sim 10^{23}$ cm$^{-2}$ within the region with density $\gtrsim 10^6$ cm$^{-3}$, we get abundances of $4\times10^{-9}$ (CH) and $9\times10^{-11}$ (CH$^+$). 

Why are the hydrides mostly in the ground-state? The excitation temperature $T_{\rm ex}$ of a species, assumed to have only two levels, is given by
\begin{equation} \label{eq:exci}
T_{\rm ex} = \frac{T_{\rm kin}}{(k T_{\rm kin}/ \Delta E) \cdot  \ln\left( 1 + \beta \ n_{\rm crit}/n({\rm H}_2) \right) \ + \ 1} \ ,
\end{equation}
with the kinetic temperature of the collision partner $T_{\rm kin}$, the energy difference between the levels $\Delta E$, the escape probability for a line photon $\beta$, the critical density $n_{\rm crit}$ and the density of a collision partner $n({\rm H}_2)$. The observed excitation temperature of CH$^+$($J=1-0$) of 38.4 K can be reproduced by Eq. \ref{eq:exci} with $\beta \sim 0.04$, assuming a kinetic temperature of 500 K, necessary for the formation of the molecule, and a H$_2$ density of order $10^6$ cm$^{-3}$, typical at scales of the \emph{Herschel} beam. This $\beta$ corresponds to an optical depth of a few at the line center, depending on the expression used for $\beta=\beta(\tau)$. Far infrared radiation by the dust continuum also pumps the molecule. Critical densities of other hydrides are not well known, but likely of similar order. Thus, the observed low excitation temperature found for OH$^+$ and H$_2$O$^+$ can be well explained by the origin of the absorption in tenuous gas.

Predictions for CH and CH$^+$ abundances and line fluxes can be made utilizing the photochemical and outflow model of \citet{Bruderer09b}. In a spherical envelope model, \citet{Bruderer10} find volume averaged abundances orders of magnitude lower than observed ($\sim6 \times 10^{-11}$ for CH and $\sim5 \times 10^{-16}$ for CH$^+$). They present also a model including a cavity etched out by the outflow, where protostellar FUV radiation can escape and directly irradiate the outflow wall. The central protostar with a luminosity of $L_{\rm bol} \sim 2 \times 10^4$ $L_\odot$ and $T_{\rm eff} \sim 3 \times 10^4$ K can heat the gas in the outflow walls to above 1000 K at scales of the \emph{Herschel} beam. Depending on the radius over which the abundance is averaged, abundances of $(0.6-2) \times 10^{-9}$ for CH and $(0.4-4) \times 10^{-9}$ for CH$^+$ are found with the outflow wall model, in acceptable agreement with the observations. 

Calculating the radiative transfer in the lines for direct comparison with observations yields fluxes of $(4.3-16.1)$ K km s$^{-1}$ for CH$^+$($J=1-0$) and $(8.9-30.8)$ K km s$^{-1}$ for CH$^+$($J=2-1$). Note that this calculation includes the influence of the pumping by far IR radiation. In the view of uncertainties (such as distance, inclination, $T_{\rm rot}$ at formation, and additional contribution to CH$^+$ by reactions of C$^+$ with vibrationally excited H$_2$; \citealt{Agundez10}), we consider the agreement with observations good. The abundances of OH$^+$ and H$_2$O$^+$ are also enhanced in the outflow walls, but lack of collisional rate coefficients prevents us from making quantitative model results.

We conclude that the scenario of FUV irradiated outflow walls can quantitatively reproduce the abundances and fluxes observed in CH and CH$^+$ emission. This indicates the presence of a large volume/mass of hot, FUV irradiated and extended gas in the envelope of this high-mass young stellar object.

\begin{table}[tb]
\caption{Fitted column density, excitation temperature, line width and position per component (see Section \ref{sec:specfit}).\label{tab:fit}}
\scriptsize
\centering
\begin{tabular}{lcllllc}
\hline
\hline
Species           & Component               & $N(X)$ & $T_{ex}$ & $\Delta$v & v$_{\rm lsr}$ & $N({\rm X})/N_{\rm H}$ \\
                  &                         & [cm$^{-2}$] & [K] & [km s$^{-1}$] & [km s$^{-1}$] & \\
\hline                                       
CH                &     \textbf{ CENTER }    & 4.0(14)    &  6.6       & 4.4     & $-5.9 $ & 4(-9)$^{b}$  \\
CH$^+$            &                          & 8.5(12)    &  38.4      & 7.0     & $-4.6 $ & 9(-11)$^{b}$ \\
\hline 
CH                &     \textbf{ RED }       & 2.6(14)    & \ldots$^a$ & 1.8     & $-0.2 $ & 2(-8)$^{c}$  \\
CH$^+$            &                          & 1.2(14)    & \ldots     & 15.0    & $ 1.6 $ & 1(-8)$^{c}$  \\
H$_2$O$^+$        &                          & 8.3(12)    & \ldots     & 4.5     & $ 2.2 $ & 7(-10)$^{c}$ \\
OH$^+$            &                          & 6.1(13)    & \ldots     & 5.1     & $ 2.8 $ & 5(-9)$^{c}$  \\
NH 974 GHz        &                          & 1.5(13)    & \ldots     & 1.5     & $-0.2 $ & 1(-9)$^{c}$  \\
NH 999 GHz        &                          & 1.5(13)    & \ldots     & 1.5     & $-0.1 $ & 1(-9)$^{c}$  \\
\hline
CH$^+$            &     \textbf{ BLUE }      & 1.8(14)    &  8.0       & 6.0     & $-17.1$ & 4(-9)$^d$    \\
H$_2$O$^+$        &                          & 5.4(12)    & \ldots     & 13.5    & $-16.4$ & 1(-10)$^d$   \\
OH$^+$            &                          & 1.6(13)    & \ldots     & 12.4    & $-16.1$ & 3(-10)$^d$   \\
NH 974 GHz        &                          & 1.9(13)    & \ldots     & 5.7     & $-12.4$ & 4(-10)$^d$   \\
NH 999 GHz        &                          & 3.8(13)    & \ldots     & 4.0     & $-10.4$ & 8(-10)$^d$   \\
\hline
\end{tabular}
\begin{flushleft} 
\footnotesize{Fractional abundances are given relative to $N_{\rm H} = N({\rm H}) + 2 N({\rm H}_2)$. $A(B) \equiv A \times 10^B$. $^a$fixed at 3 K.\ $^b$ using $N_{\rm H} \sim 10^{23}$ cm$^{-2}$, see text.\\
$^c$ using $N_{\rm H} \sim 1.2 \times 10^{22}$ cm$^{-2}$ (\citealt{Minh08}).\\
$^d$ using $N_{\rm H} \sim 4.6 \times 10^{22}$ cm$^{-2}$ (\citealt{Mitchell89}).}
\end{flushleft}
\end{table}

\section{Conclusion}  \label{sec:conclusion}

We use the new \emph{Herschel}-HIFI instrument to observe light hydrides towards the high-mass star forming region AFGL 2591. Besides OH$^+$, H$_2$O$^+$ and NH detected in absorption only, we see CH and CH$^+$ in emission. The red and blue velocity shifted absorption can be assigned to a foreground cloud and an outflow lobe directed towards us, respectively. The emission of CH and CH$^+$ is at the source velocity. Its abundance and excitation can be explained in the scenario of direct FUV irradiation of the envelope through a low density cavity of the outflow region.

\bibliographystyle{aa}

\begin{thebibliography}{32}
\expandafter\ifx\csname natexlab\endcsname\relax\def\natexlab#1{#1}\fi

\bibitem[{{Ag{\'u}ndez} {et~al.}(2010){Ag{\'u}ndez}, {Goicoechea},
  {Cernicharo}, {Faure}, \& {Roueff}}]{Agundez10}
{Ag{\'u}ndez}, M., {Goicoechea}, J.~R., {Cernicharo}, J., {Faure}, A., \&
  {Roueff}, E. 2010, \apj, 713, 662

\bibitem[{{Benz} {et~al.}(2010)}]{Benz10}
{Benz}, A.~O., {Bruderer}, S., {van Dishoeck}, E.~F., {et~al.} 2010, \aap, this volume

\bibitem[{{Black}(1998)}]{Black98}
{Black}, J.~H. 1998, Faraday Discussions No. 109, 257

\bibitem[{{Bruderer} {et~al.}(2009{\natexlab{a}}){Bruderer}, {Benz}, {Bourke},
  \& {Doty}}]{Bruderer09c}
{Bruderer}, S., {Benz}, A.~O., {Bourke}, T.~L., \& {Doty}, S.~D.
  2009{\natexlab{a}}, \aap, 503, L13

\bibitem[{{Bruderer} {et~al.}(2009{\natexlab{b}}){Bruderer}, {Benz}, {Doty},
  {van Dishoeck}, \& {Bourke}}]{Bruderer09b}
{Bruderer}, S., {Benz}, A.~O., {Doty}, S.~D., {van Dishoeck}, E.~F., \&
  {Bourke}, T.~L. 2009{\natexlab{b}}, \apj, 700, 872

\bibitem[{{Bruderer} {et~al.}(2010){Bruderer}, {Benz}, {St\"auber}, \&
  {Doty}}]{Bruderer10}
{Bruderer}, S., {Benz}, A.~O., {St\"auber}, P., \& {Doty}, S.~D. 2010, \apj,
  in press

\bibitem[{{de Graauw} {et~al.}(2010)}]{deGraauw10}
{de Graauw}, T., {Helmich}, F.~P., {Phillips}, T.~G., {et~al.} 2010, \aap, 518, L6

\bibitem[{{Doty} {et~al.}(2002){Doty}, {van Dishoeck}, {van der Tak}, \&
  {Boonman}}]{Doty02}
{Doty}, S.~D., {van Dishoeck}, E.~F., {van der Tak}, F.~F.~S., \& {Boonman},
  A.~M.~S. 2002, \aap, 389, 446

\bibitem[{{Gerin} {et~al.}(2010)}]{Gerin10}
{Gerin}, M., {De Luca}, M., {Black}, J., {et~al.} 2010, \aap, 518, L110

\bibitem[{{Godard} {et~al.}(2009){Godard}, {Falgarone}, \& {Pineau Des
  For{\^e}ts}}]{Godard09}
{Godard}, B., {Falgarone}, E., \& {Pineau Des For{\^e}ts}, G. 2009, \aap, 495,
  847

\bibitem[{{Hammami} {et~al.}(2009){Hammami}, {Owono Owono}, \&
  {St{\"a}uber}}]{Hammami09}
{Hammami}, K., {Owono Owono}, L.~C., \& {St{\"a}uber}, P. 2009, \aap, 507, 1083

\bibitem[{{Hasegawa} \& {Mitchell}(1995)}]{Hasegawa95}
{Hasegawa}, T.~I. \& {Mitchell}, G.~F. 1995, \apj, 451, 225

\bibitem[{{H{\"u}bers} {et~al.}(2009){H{\"u}bers}, {Evenson}, {Hill}, \&
  {Brown}}]{Huebers09}
{H{\"u}bers}, H., {Evenson}, K.~M., {Hill}, C., \& {Brown}, J.~M. 2009, \jcp,
  131, 034311

\bibitem[{{Le Petit} {et~al.}(2004){Le Petit}, {Roueff}, \&
  {Herbst}}]{LePetit04}
{Le Petit}, F., {Roueff}, E., \& {Herbst}, E. 2004, \aap, 417, 993

\bibitem[{{Menten} {et~al.}(2010){Menten}, {Wyrowski}, {Belloche},
  {et~al.}}]{Menten10}
{Menten}, K., {Wyrowski}, F., {Belloche}, A., {et~al.} 2010, \aap, submitted

\bibitem[{{Minh} \& {Yang}(2008)}]{Minh08}
{Minh}, Y.~C. \& {Yang}, J. 2008, Journal of Korean Astronomical Society, 41,
  139

\bibitem[{{Mitchell} {et~al.}(1989){Mitchell}, {Curry}, {Maillard}, \&
  {Allen}}]{Mitchell89}
{Mitchell}, G.~F., {Curry}, C., {Maillard}, J., \& {Allen}, M. 1989, \apj, 341,
  1020

\bibitem[{{Mitchell} {et~al.}(1992){Mitchell}, {Hasegawa}, \&
  {Schella}}]{Mitchell92}
{Mitchell}, G.~F., {Hasegawa}, T.~I., \& {Schella}, J. 1992, \apj, 386, 604

\bibitem[{{M{\"u}ller} {et~al.}(2005){M{\"u}ller}, {Schl{\"o}der}, {Stutzki},
  \& {Winnewisser}}]{Mueller05}
{M{\"u}ller}, H.~S.~P., {Schl{\"o}der}, F., {Stutzki}, J., \& {Winnewisser}, G.
  2005, Journal of Molecular Structure, 742, 215

\bibitem[{{M\"urtz} {et~al.}(1998){M\"urtz}, {Zink}, {Evenson}, \&
  {Brown}}]{Muertz98}
{M\"urtz}, P., {Zink}, L.~R., {Evenson}, K.~M., \& {Brown}, J.~M. 1998, J.
  Chem. Phys., 109, 9744

\bibitem[{{Ossenkopf} {et~al.}(2010)}]{Ossenkopf10}
{Ossenkopf}, V., {M\"uller}, H.~S.~P., {Lis}, D.~C. {et~al.} 2010, \aap, 518, L111

\bibitem[{{Preibisch} {et~al.}(2003){Preibisch}, {Balega}, {Schertl}, \&
  {Weigelt}}]{Preibisch03}
{Preibisch}, T., {Balega}, Y.~Y., {Schertl}, D., \& {Weigelt}, G. 2003, \aap,
  412, 735

\bibitem[{{St{\"a}uber} \& {Bruderer}(2009)}]{Staeuber09}
{St{\"a}uber}, P. \& {Bruderer}, S. 2009, \aap, 505, 195

\bibitem[{{Sternberg} \& {Dalgarno}(1995)}]{Sternberg95}
{Sternberg}, A. \& {Dalgarno}, A. 1995, \apjs, 99, 565

\bibitem[{{van der Tak} {et~al.}(2003){van der Tak}, {Boonman}, {Braakman}, \&
  {van Dishoeck}}]{vdTak03}
{van der Tak}, F.~F.~S., {Boonman}, A.~M.~S., {Braakman}, R., \& {van
  Dishoeck}, E.~F. 2003, \aap, 412, 133

\bibitem[{{van der Tak} {et~al.}(1999){van der Tak}, {van Dishoeck}, {Evans},
  {Bakker}, \& {Blake}}]{vdTak99}
{van der Tak}, F.~F.~S., {van Dishoeck}, E.~F., {Evans}, II, N.~J., {Bakker},
  E.~J., \& {Blake}, G.~A. 1999, \apj, 522, 991

\bibitem[{{Wilson} \& {Rood}(1994)}]{Wilson94}
{Wilson}, T.~L. \& {Rood}, R. 1994, \araa, 32, 191

\end{thebibliography}

\begin{acknowledgements}
We thank the anonymous referee for useful comments. The work on star formation at ETH Zurich is partially funded by the Swiss National Science Foundation grant 200020-113556. This program is made possible thanks to the Swiss HIFI guaranteed time program. HIFI has been designed and built by a consortium of institutes and university departments from acrossEurope, Canada and the United States under the leadership of SRON Netherlands Institute for Space Research, Groningen, The Netherlands and with major contributions from Germany, France and the US.Consortium members are: Canada: CSA, U.~Waterloo; France: CESR, LAB, LERMA, IRAM; Germany:KOSMA, MPIfR, MPS; Ireland, NUI Maynooth; Italy: ASI, IFSI-INAF, Osservatorio Astrofisico di Arcetri-INAF; Netherlands: SRON, TUD; Poland: CAMK, CBK; Spain: Observatorio Astron\'omico Nacional (IGN),Centro de Astrobiolog\'a (CSIC-INTA). Sweden: Chalmers University of Technology - MC2, RSS \& GARD;Onsala Space Observatory; Swedish National Space Board, Stockholm University - Stockholm Observatory;Switzerland: ETH Zurich, FHNW; USA: Caltech, JPL, NHSC. HIPE is a joint development by the Herschel Science Ground Segment Consortium, consisting of ESA, the NASA Herschel Science Center, and the HIFI, PACS and SPIRE consortia.
\end{acknowledgements}

\Online 

\begin{appendix}

\section{Spectra of non-detections} \label{sec:nondetect}

\begin{figure}[htb]
\includegraphics[width=\hsize]{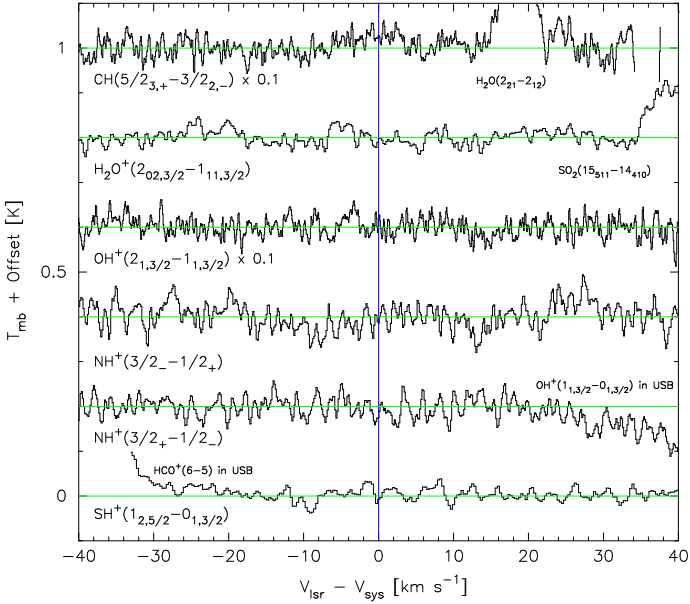}
\caption{Spectra observed towards AFGL 2591 without detection of target lines. The velocity is given relative to the systemic velocity v$_{\rm sys} = -5.5$ km s$^{-1}$.}
\label{fig:spectranon}
\end{figure}

Figure \ref{fig:spectranon} presents spectra without detected target lines, given in Table \ref{tab:lines}. For clarity, the spectra of CH($5/2_{3,+} - 3/2_{2,-}$) and OH$^+$($2_{1,3/2}-1_{1,3/2}$) have been scaled by a factor of 0.1. In the spectrum of CH($5/2_{3,+} - 3/2_{2,-}$), the H$_2$O($2_{21}-2_{12}$) line is detected, which will be presented in an upcoming paper (van der Tak et al., in preparation). The H$_2$O$^+$($2_{02,3/2}-1_{11,3/2}$) spectrum contains a SO$_2$ line with upper level energy similar to the lines detected by \citet{vdTak03} in the same source. In the other sideband of the NH$^+$($3/2_+ -1/2_-$) spectrum, the OH$^+$($2_{1,3/2}-1_{1,3/2}$) line is seen. In the upper side band of the SH$^+$($1_{2,5/2}-0_{1,3/2}$) line, the HCO$^+$($6-5$) line has been detected. The velocity range of $\pm 40$ km s$^{-1}$ corresponds to a frequency range of $\pm 67$ MHz at 500 GHz and $\pm 253$ MHz at 1900 GHz. Thus, the lines should be seen in the spectra, even if the predicted rest frequency is grossly wrong.

\section{Modeling the line spectra} \label{sec:rtsolution}

This appendix gives the equations used to model the line spectra. The intensity $T^{\rm model}(\nu)$ [K] at frequency $\nu$ for a slab of $N$ emitting or absorbing layers in front of a continuum source is obtained from
\begin{eqnarray}
T^{\rm model}(\nu)&=&T^{\rm continuum}(\nu) e^{-\sum_{i=1}^N \Delta \tau^i(\nu)}  \\
&&+  \sum_{i=1}^N \frac{c^2}{2 \nu_0^2 k} B_{\nu_0}(T^i_{\rm ex}) \ e^{-\sum_{j=1}^{i-1} \Delta \tau^j(\nu)} \left(1-e^{-\Delta \tau^i(\nu)} \right) \ , \nonumber
\end{eqnarray}
with the (background) continuum intensity $T^{\rm cont}(\nu)$, the opacity per layer $\Delta \tau^i(\nu)$, the frequency of the  component $\nu_0$ with the largest Einstein-A coefficient and the Planck function $B_\nu(T)$. Note that this equation assumes a covering factor of 100 \% for each layer and the same excitation temperature for all hyperfine/fine components. The line opacity is calculated from the sum over the $M$ fine/hyperfine components of the transition,
\begin{equation}
\tau^i(\nu)=N_{\rm mol}^i \ \sum_{k=1}^M \ \frac{c^2}{8 \pi (\nu^k)^2} \ A_{ul}^k \ \left(x^{i,k}_l \frac{g_u}{g_l} - x^{i,k}_u \right) \ \phi^{i,k}(\nu) \ ,
\end{equation}
with the column density per layer $N_{\rm mol}^i$, the Einstein-A coefficient $A_{ul}^k$, the normalized line profile function $\phi^{i,k}(\nu)$, the normalized level population of the upper and lower level ($x^{i,k}_u$ and $x^{i,k}_l$, respectively) and the statistical weights ($g_u$ and $g_l$, respectively). The normalized level population per layer is a Boltzmann distribution for the temperature $T^i_{\rm ex}$. The line profile function $\phi^{i,k}(\nu)$ is assumed to be a Gaussian with width $\Delta$v$^i$ (FWHM) centered at the frequency of the component minus the velocity v$_{\rm lsr}^i$ of the layer.

The free parameters per layer are thus the excitation temperature $T^i_{\rm ex}$, the column density $N_{\rm mol}^i$, the width $\Delta$v$^i$ and the velocity v$_{\rm lsr}^i$. Two layers are used for CH, H$_2$O$^+$, OH$^+$ and NH with a red-shifted layer in front of a blue-shifted layer. Three layers are used for CH$^+$ with a layer centered at the source velocity between the background continuum and the blue-shifted layer. To constrain the parameters of the layer, the least squares of the modeled to the observed spectra are minimized.

Molecular data of the individual hyperfine components of CH, OH$^+$, NH and H$_2$O$^+$ used to model the line spectra in Figure \ref{fig:spectra} are given in Table \ref{tab:hyper}. The table gives the frequency, Einstein-A coefficient ($A_{ul}$), the upper level energy ($E_u$) and the statistical weights of the upper and lower level of the transition ($g_u$ and $g_l$, respectively). The third column refers to the velocity shift in km s$^{-1}$ of the component relative to the one with the largest Einstein-A coefficient. The component with the largest Einstein-A coefficient is marked by (*).

\begin{table}[htb]
\caption{Hyperfine components used to model the spectra.\label{tab:hyper}}
\centering
\begin{tabular}{lcccccc}
\hline
\hline
Species       & Frequency  & Shift to (*)  & $E_{u}$ &$A_{ul}$    & $g_u$ & $g_l$ \\
              & [GHz]      & [km s$^{-1}$] & [K]     & [s$^{-1}$] &       &       \\
\hline
CH (*)        &   536.7611 &        0.0 &      25.76 &    6.4(-4) &    5 &    3 \\
CH            &   536.7820 &      -11.6 &      25.76 &    2.1(-4) &    3 &    3 \\
CH            &   536.7957 &      -19.3 &      25.76 &    4.3(-4) &    3 &    1 \\
\hline
H$_2$O$^+$    &  1115.1505 &       14.4 &      53.51 &    1.7(-2) &    4 &    2 \\
H$_2$O$^+$    &  1115.1862 &        4.8 &      53.51 &    2.8(-2) &    2 &    2 \\
H$_2$O$^+$ (*)&  1115.2041 &        0.0 &      53.51 &    3.1(-2) &    6 &    4 \\
H$_2$O$^+$    &  1115.2632 &      -15.9 &      53.51 &    1.4(-2) &    4 &    4 \\
H$_2$O$^+$    &  1115.2989 &      -25.5 &      53.51 &    3.5(-3) &    2 &    4 \\
\hline 
OH$^+$        &  1032.9979 &       35.0 &      49.58 &    1.4(-2) &    2 &    2 \\
OH$^+$        &  1033.0044 &       33.1 &      49.58 &    3.5(-3) &    4 &    2 \\
OH$^+$        &  1033.1118 &        2.0 &      49.58 &    7.0(-3) &    2 &    4 \\
OH$^+$ (*)    &  1033.1186 &        0.0 &      49.58 &    1.8(-2) &    4 &    4 \\
\hline
NH            &   974.3156 &       50.1 &      46.77 &    1.8(-6) &    4 &    4 \\
NH            &   974.3426 &       41.8 &      46.77 &    6.0(-6) &    6 &    4 \\
NH            &   974.3546 &       38.1 &      46.77 &    4.5(-6) &    4 &    2 \\
NH            &   974.4106 &       20.9 &      46.77 &    8.9(-5) &    4 &    6 \\
NH            &   974.4114 &       20.6 &      46.77 &    5.1(-4) &    2 &    4 \\
NH            &   974.4363 &       12.9 &      46.77 &    2.3(-3) &    4 &    4 \\
NH            &   974.4375 &       12.6 &      46.77 &    1.3(-3) &    6 &    6 \\
NH            &   974.4440 &       10.6 &      46.77 &    1.8(-3) &    4 &    4 \\
NH            &   974.4504 &        8.6 &      46.77 &    4.8(-3) &    2 &    2 \\
NH            &   974.4622 &        5.0 &      46.77 &    5.0(-3) &    4 &    2 \\
NH            &   974.4710 &        2.3 &      46.77 &    5.7(-3) &    6 &    4 \\
NH            &   974.4754 &        0.9 &      46.77 &    3.4(-3) &    4 &    2 \\
NH (*)        &   974.4784 &        0.0 &      46.77 &    6.9(-3) &    8 &    6 \\
NH            &   974.4793 &       -0.3 &      46.77 &    6.0(-3) &    6 &    4 \\
NH            &   974.5313 &      -16.3 &      46.77 &    2.6(-4) &    4 &    6 \\
NH            &   974.5398 &      -18.9 &      46.77 &    6.5(-4) &    2 &    4 \\
NH            &   974.5581 &      -24.5 &      46.77 &    9.8(-4) &    2 &    2 \\
NH            &   974.5648 &      -26.6 &      46.77 &    7.7(-4) &    4 &    4 \\
NH            &   974.5744 &      -29.5 &      46.77 &    8.1(-4) &    6 &    6 \\
NH            &   974.5830 &      -32.2 &      46.77 &    2.6(-4) &    4 &    2 \\
NH            &   974.6078 &      -39.8 &      46.77 &    1.2(-4) &    6 &    4 \\
\hline
NH            &   999.8784 &       28.5 &      47.99 &    9.9(-4) &    6 &    4 \\
NH            &   999.8821 &       27.4 &      47.99 &    3.9(-3) &    4 &    4 \\
NH            &   999.9060 &       20.2 &      47.99 &    3.1(-7) &    4 &    4 \\
NH            &   999.9079 &       19.6 &      47.99 &    3.0(-3) &    2 &    4 \\
NH            &   999.9192 &       16.3 &      48.00 &    1.1(-3) &    2 &    4 \\
NH            &   999.9212 &       15.7 &      47.99 &    9.6(-4) &    4 &    2 \\
NH            &   999.9451 &        8.5 &      47.99 &    1.8(-3) &    4 &    2 \\
NH            &   999.9470 &        7.9 &      47.99 &    1.8(-4) &    2 &    2 \\
NH            &   999.9582 &        4.5 &      48.00 &    1.7(-3) &    2 &    2 \\
NH (*)        &   999.9734 &        0.0 &      47.99 &    5.2(-3) &    6 &    6 \\
NH            &   999.9771 &       -1.1 &      47.99 &    3.9(-4) &    4 &    6 \\
NH            &  1000.0010 &       -8.3 &      47.99 &    2.9(-3) &    4 &    6 \\
NH            &  1000.0068 &      -10.0 &      47.99 &    1.2(-3) &    6 &    4 \\
NH            &  1000.0105 &      -11.1 &      47.99 &    1.7(-3) &    4 &    4 \\
NH            &  1000.0288 &      -16.6 &      47.99 &    4.9(-4) &    4 &    2 \\
NH            &  1000.0345 &      -18.3 &      47.99 &    1.6(-3) &    4 &    4 \\
NH            &  1000.0363 &      -18.9 &      47.99 &    4.4(-5) &    2 &    4 \\
NH            &  1000.0476 &      -22.2 &      48.00 &    4.5(-3) &    2 &    4 \\
NH            &  1000.0527 &      -23.8 &      47.99 &    1.1(-3) &    4 &    2 \\
NH            &  1000.0546 &      -24.3 &      47.99 &    4.1(-3) &    2 &    2 \\
NH            &  1000.0659 &      -27.7 &      48.00 &    1.6(-4) &    2 &    2 \\
\hline
\end{tabular}
\begin{flushleft}
\footnotesize{$A(B) \equiv A \times 10^B$.}
\end{flushleft}
\end{table}

\end{appendix}

\end{document}